\documentclass[%
,preprint%
,secnumarabic%7
,amssymb]{revtex4}
\usepackage{amsmath}%
\usepackage{bm}%
\usepackage{epsf}%
\newif\ifpdf
\ifx\pdfoutput\undefined
\pdffalse % we are not running PDFLaTeX
\else
\pdfoutput=1 % we are running PDFLaTeX
\pdftrue
\fi
\ifpdf
\usepackage[pdftex]{graphicx}
\else
\usepackage{graphicx}
\fi

\ifpdf
\DeclareGraphicsExtensions{.pdf, .jpg}
\else
\DeclareGraphicsExtensions{.eps, .jpg}
\fi
\begin{document}
\title{Differential probability for surface and volume electronic excitations in Fe, Pd and Pt.}

\author{ Wolfgang S.M. Werner\footnote{werner@iap.tuwien.ac.at, \\
fax:+43-1-58801-13499,~tel:+43-1-58801-13462}}%
\affiliation{Institut f{\" u}r Allgemeine Physik,~Vienna University of Technology,
Wiedner Hauptstra\ss e 8--10,~A~1040~Vienna,~Austria
}
\date{\today}%
%\tableofcontents
%\twocolumn
\begin{abstract} 
The normalized differential mean free path for volume scattering and the differential surface excitation probability
for medium energy electrons travelling in Fe, Pd and Pt are extracted from Reflection Electron Energy Loss
Spectra (REELS). This was achieved by means of a recently introduced procedure in which two REELS spectra taken under
different experimental conditions are simultaneously deconvoluted. In this way, it is possible  to obtain the unique
reconstruction for the surface and volume single scattering loss distribution. The employed method is compared with a  procedure that is
frequently used for this purpose [Tougaard and Chorkendorff, Phys. Rev. B 35(1987)6570]. It is shown, both
theoretically and through analysis of model spectra as well as experimental data that this method does not result
in a {\em single} scattering loss distribution. Rather, it gives a mixture of surface,
bulk and mixed scattering of any order. 
%These findings explain the spurious negative excursions in the
%retrieved loss distributions that have been published over the past twenty years.
\end{abstract}

\maketitle 

\section{Introduction}
Quantitative analysis of reflection electron energy loss spectra (REELS) is important for surface science both in a
practical and fundamental sense since such spectra contain valuable information on the response of a solid to an
external electromagnetic perturbation. From the fundamental point of view it is  desirable to extract
 the frequency ($\omega$) and momentum ($q$) dependent dielectric function $\varepsilon(q,\omega)$ from experimental data, being
the characteristic quantity for the susceptibility of the solid to polarize under bombardment with electromagnetic 
radiation
\cite{palik,henke,chantler,ambroschpr,bechjphys}.
 For practical applications of electron spectroscopy, information about the differential distribution of energy losses
in an individual loss process is a  prerequisite for quantitative analysis of electron spectra
\cite{toupr35,werqsasia}. Such data can be directly extracted from REELS spectra and converted to dielectric data by invoking linear response theory.

The interaction of  a reflected electron with a solid is quite complex since different types of collisions occur along
the electron trajectory, and, furthermore, these scattering processes can occur repeatedly when the
pathlength of the electron in the solid is long compared to the relevant mean free paths for scattering. The changes in
the degrees of freedom of the electron are brought about by the interaction with the ionic subsystem, which mainly
changes the direction of the particle in an elastic process and the interaction with the electronic subsystem, that gives
rise to energy losses \cite{werqsasia}. An energy loss process  can be  conceived as a decelaration of the
probe electron by a polarization field  that is set up in the solid by the response of the weakly--bound solid state
electrons to the incoming electron \cite{landauem}. This physical picture provides the  connection between the dielectric function and 
the differential mean free path for inelastic electron scattering. Near a surface, the situation is further complicated
through the occurence of so--called surface excitations
\cite{ritchiepr106,sternpr120,echeniquependry,echenique,yubpr,aripr49,tungpr49,dentpra57,vicanek,aminovpr63}. 
 These are additional modes of the inelastic interaction that
result from the boundary conditions of Maxwell's equations near an interface between regions of  different
electromagnetic susceptibility. Surface excitations have a lower resonance frequency as compared to volume excitations
and they occur on both sides of the solid-vacuum interface, but the probability for a surface excitation decays
rapidly with the distance from the interface.

The question then arises how to extract the single scattering loss distribution for volume and surface excitations
from an experimental spectrum. An important step to resolve this problem was taken  some time ago by the pioneering work of Tougaard and
Chorkendorff \cite{toupr35} who found a simple deconvolution formula for REELS spectra. However, the physical
model on which it is based completely ignores  surface excitations. Furthermore, the basic process in a
reflection experiment, multiple elastic scattering, was treated  approximately, in the so--called P$_1$-approximation, in which the 
expansion of the elastic scattering cross section in spherical harmonics is terminated after the first order
\cite{davison}. In the medium energy range, the higher order phase shifts are by no means negligible
\cite{werhayreels} and consequently the P$_1$--approximation does not provide a good physical basis to model REELS
spectra.

 Nonetheless, this formula is extensively used nowadays and
it is commonly  believed that the "effective" cross section that is extracted from a REELS spectrum in this way indeed represents a
single  scattering loss distribution, although the exact nature of this "effective" cross section has never been
analyzed.

Some years later, several authors independently  introduced a deconvolution procedure for REELS spectra in which elastic
scattering is treated more realistically \cite{werpia,vicanek}. Although this provides a better model to describe REELS
spectra, the resulting deconvolution procedure leads to results that are very similar to those using  Tougaard
and Chorkendorffs  algorithm. This has been attributed to the fact that surface excitations are also neglected in this
procedure \cite{werpia,vicanek,chenprb58}.

Recently, a deconvolution procedure was introduced that is based on a physical  model that  takes into account
the multiple occurence of surface excitations and deflections during elastic processes \cite{werbivar}. It
is based on the simultaneous deconvolution of a set of two REELS spectra taken at different energies or geometrical
configurations for which the relative contribution of surface and bulk excitations differs substantially.
The main assumption in this procedure is that the {\em normalized} single scattering loss distributions for surface
and volume scattering are independent of the energy of the probing electron and the angle of surface crossing
\cite{werbivar,werqsasia}. 
 The essential difference with the earlier algorithms is that the latter are essentially effected by reversion of a
univariate power series in Fourier space which has no unique solution if more than one kind of inelastic proces takes
place. The algorithm of Ref.\cite{werbivar} is based on reversion of a {\em bivariate} power series in Fourier space,
which does provide the unique solution that is straightforward to interpret quantitatively. 

The procedure based on a bivariate power series reversion was succesfully applied to REELS spectra of Si, Cu and
Au. The deconvolution procedure is
quite involved, however, and the question therefore arises how seriously the deficiencies of the earlier methods
addressed above affect the results of the deconvolution.
The present paper addresses this issue. The theory of signal generation in a REELS experiment as well as the different
deconvolution procedures are reviewed.  The reason for the good performance of the P$_1$-approximation for
{\em deconvoluting} spectra --in spite of the fact that it is not a good model to {\em simulate} REELS spectra-- is
elucidated.  It is shown theoretically that the procedure of Tougaard and Chorkendorff does  not  result in a
single scattering loss distribution. Rather, this procedure provides a mixture of
surface and  volume  multiple scattering loss distributions and a contribution of mixed terms of any scattering
order. Therefore, quantitative interpretation of the resulting loss distributions is troublesome.

Results of analysis of experimental spectra for Fe, Pd and Pt using a bivariate power series reversion are compared
with results of the earlier algorithms. The bivariate power series reversion leads to true single scattering
loss distributions that agree quantitatively with results based on the theory given by Tung \cite{tungpr49} on an
absolute scale.
%, the univariate power series reversion does not lead to an "effective"  loss distribution that can be
%represented by a linear combination of the single scattering distributions for surface and volume scattering.
The main reason for the essential failure of any method based on a univariate power series reversion is that it does not eliminate mixed
terms corresponding to elecrons that experience both surface and volume excitations a certain number of times.

\section{Theoretical}
The theory of signal generation in REELS is discussed in detail in Ref~\cite{werbivar}, here we only present a brief synopsis.
The flux of medium-energy  electrons reflected from a solid surface consists of particles that have travelled
different pathlengths in the solid. Since the probability for (multiple) scattering increases monotonically with the
travelled pathlength, the outgoing flux consists of groups of electrons that have experienced a different number of
inelastic processes. The net energy loss experienced after a given number of inelastic processes is subject to
fluctuations that can be expressed in terms of a multiple self-convolution of the single scattering energy fluctuation
distributions as follows from solution of a Boltzmann-type kinetic equation \cite{werepes3,werschatteels}. The single scattering
fluctuation distributions are given by the inverse differential mean free path for scattering.
The normalized
inverse differential mean free path for surface (subscript "s") and bulk (subscript "b") excitations will be denoted by
$w_s(T)$
and $w_b(T)$ in the following. The normalized bulk mean free path is related to the absolute differential mean free path
$W_b(T,E)$ via the
total (integrated) inelastic mean free path  (IMFP) $\lambda_i(E)$:
\begin{equation}
\label{enormbulk}
w_b(T,E)=\lambda_i (E) W_b(T,E)
\end{equation}
The normalized single scattering surface loss distribution is given in terms of the differential mean free path
integrated over the part of the trajectory in the surface scattering zone, $W_s(T,E,\theta)$, as:
\begin{equation}
\label{enormsurf}
w_s(T,E.\theta)=\frac{ W_s(T,E,\theta)}{\langle n_s(E,\theta) \rangle}
\end{equation}
where the (dimensionless) quantity $\langle n_s  (E,\theta) \rangle$ represents the average number of surface excitations in a
single surface crossing along the polar direction $\cos\theta$. The dimensions of $w_s(T,E,\theta)$ and $w_b(T,E)$ are the same, reciprocal eV, while the dimensions of
the quantities $W_s(T,E,\theta)$ and $W_b(T,E)$ are different, as is their physical meaning (see Ref.~\cite{werbivar}). 
It turns out \cite{werqsasia,wernrelprl,werbivar} that the {\em shape} of the single scattering loss distributions depends only very weakly on the energy and, 
in the case of surface excitations, on the surface crossing direction. Then the normalized single scattering loss
distributions are only a function of the 
energy loss to a good approximation:
\begin{eqnarray}
\label{enormdiimfpdsep}
w_{b}(T,E)&\simeq& w_{b}(T)\nonumber\\
w_{s}(T,E,\theta)&\simeq& w_{s}(T)
\end{eqnarray} 
Formulae
to establish the differential inverse inelastic mean free path (DIIMFP) $W_b(T,E)$ and the differential surface
excitation probability (DSEP)
$W_s(T,E,\theta)$ can be found in
many instances in the literature, in the present work the expressions given in Ref.~\cite{tungpr49} will be employed.
 The differential mean free path for inelastic scattering   is frequently 
refered to in the literature  as "cross section" but one should keep in mind that in doing so, the solid state character of the medium is
completely ignored, which is certainly not justifiable for medium energy electrons travelling in solids.

In the following, the fluctuation distributions after $n_s$-fold surface and $n_b$-fold volume scattering will be
denoted by $w_s^{(n_s)}$ and $w_b^{(n_b)}$, respectively, where the exponent in brackets ($^{(n)}$) indicate the
$n-1$--fold selfconvolution of the respective quantity. Multiplying these with the relative number of electrons that
arrive in the detector after suffering the corresponding number of energy losses, the so--called
partial intensities, $\alpha_{n_b,n_s}$, and summing over all scattering orders, one obtains the following expression for the loss spectrum
$y_L(T)$:
\begin{equation}
\label{eyieldloss}
y_L(T)=
\sum\limits_{n_b=0}^\infty
\sum\limits_{n_s=0}^\infty
\alpha_{n_b, n_s}
w_b^{(n_b)}(T')
\otimes
w_s^{(n_s)}(T-T')
\end{equation}
where the symbol $\otimes$ denotes a convolution over the energy loss $T$. The spectrum $y_L(T)$ represents the loss
spectrum divided by the area  of the elastic peak and from which the elastic peak is removed by choosing $\alpha_{n_b=0,
n_s=0}=0$. 

The reduced partial intensities for surface and bulk scattering are uncorrelated:
\begin{equation}
\label{epiuncorr}
\alpha_{n_b, n_s}=  \alpha_{n_b}\times \alpha_{n_s}
\end{equation}
since the part of the trajectory in the surface scattering zone is rectilinear to a good approximation \cite{wernrelprl}.
Although (small) effects of  deflections in the surface scattering zone have been experimentally observed for a rather
pathological case \cite{werzemsep}, this approach nonetheless has proven to constitute an effective approximation
\cite{werzemsep,wernrelprl,wernrelsia,wercoissl}.

The reduced bulk partial intensities are given by the well known equation \cite{werqsasia,werprcsd}
\begin{equation}
\label{epipld}
\alpha_{n_b}=
\frac
{\int\limits_0^\infty  W_{n_b}(s) Q(s) ds}
{\int\limits_0^\infty  W_{0}(s) Q(s) ds}
\end{equation}
Here $Q(s)$ is the distribution of pathlengths and $W_{n}(s)$ is the stochastic process for multiple scattering, which,
in the quasielastic regime, is given by \cite{werprcsd}:
\begin{equation}
\label{epoisson}
W_{n}(s)=\Big(\frac{s}{\lambda_i}\Big)^{n}\frac{e^{-s/\lambda_i}}{n!}
\end{equation}

The partial intensities for surface scattering, $A_{n_{s}}$, follow Poisson statistics, implying that the reduced partial intensities
$\alpha_{n_{s}}=A_{n_{s}}/A_{n_{s}=0}$ are given by:
\begin{equation}
\label{episurf}
\alpha_{n_s}=\frac{\langle n_s \rangle^{n_s}}{n_s!}
\end{equation}
 with $ \langle n_s\rangle=\langle n_s^i\rangle+\langle n_s^o\rangle$ representing the sum of the
average number of surface excitations along the in and outgoing part of the trajectory. Surface excitations along the in- and outgoing part of the trajectory are combined in Eqn.~(\ref{eyieldloss})
by assuming that the shape of the {\em normalized}Êdistribution of surface energy losses
is independent of the surface crossing direction $\theta$ \cite{werbivar}.

%, while the number of surface excitations is of course dependent on
%the surface crossing direction. In the present work, the following expression is used to estimate the average number
%of surface excitations in a single surface crossing \cite{wersuxssl}:
%\begin{equation}
%\label{esepemp}
%\langle n_s(E,\theta) \rangle =\frac{1}{a_s\sqrt{E}\cos\theta+1}
%\end{equation}
%The quantity $a_s$ is given for a large number of elemental solids in Ref.~\cite{wersuxssl} in units of the free electron value $a_{NFE}=\sqrt{8a_0/\pi^2e^2}=0.173
%eV^{-1/2}$, where   $e^2=14.4eV\AA$ is the elementary charge squared and  $a_0=0.52\AA$ is the Bohr radius.
%An empirical relationship between the surface excitation parameter $a_s$ and the generalized plasmon energy was also
%derived, allowing one to estimate the extent of surface excitations for an arbitrary material.

\section{Decomposition of REELS spectra}
\subsection{ I: reversion of a univariate power series within the P$_1$--approximation (Ref.~\cite{toupr35}) .} 
In the method of Tougaard and Chorkendorf \cite{toupr35} two approximations are made: (1) surface excitations are
neglected; and (2) multiple elastic scattering is treated in the so--called P$_1$--approximation. Based on
these starting assumptions, a simple deconvolution formula was derived that is frequently used also in situations were
surface excitations dominate the low energy loss region of a  spectrum. It is commonly
believed that in such cases the method  provides an "effective" {\em single} scattering loss distribution i.e. some
kind of linear combination of $w_s(T)$ and $w_b(T)$ but  this has never been proved.

In the P$_1$--approximation, the elastic scattering cross section that enters the collision term of a Boltzmann-type
kinetic equation for the particle transfer, is expanded in spherical harmonics and the expansion is terminated after
the first order. In this way, a tractable analytic solution for the reflection problem has been found \cite{tofpr32}.
The expression for the pathlength distribution  $Q_{P_1}(s)$ reads:
\begin{equation}
\label{epld}
Q_{P_1}(s)\propto \exp   (-s/L)
\end{equation}
where the characteristic length $L$ is about twice the transport mean free path $L\simeq 2\lambda_{tr}$. Inserting
this result into Equation~(\ref{epipld}), the bulk partial intensities are found as:
\begin{equation}
\label{epip1}
\alpha_{n_b}^{P_1}=\kappa^{n_b}
\end{equation}
where $\kappa=L/(L+\lambda_i)$.

When surface excitations are neglected, Equation~(\ref{eyieldloss}) simplifies to:
\begin{equation}
\label{eyieldunivariate}
y_L(T)=
\sum\limits_{n_b=0}^\infty
\alpha_{n_b}
w_b^{(n_b)}(T)
\end{equation}
Convoluting this equation with $\alpha_2/\alpha_1w_b(T)$ and subtracting the result from
Equation~(\ref{eyieldunivariate}) gives:
\begin{equation}
\label{etmp1}
y_L(T)-\frac{\alpha_2}{\alpha_1}w_b(T')\otimes y_L(T-T')=\alpha_1 w_{b}(T)+\sum\limits_{n=3}^\infty
(\alpha_n-\frac{\alpha_2}{\alpha_1}\alpha_{n-1})w_b^{(n)}
\end{equation}
For a sequence of partial intensities of the form $\alpha_{n}=\kappa^{n}$, all terms in the sum on the right
hand side vanish. Therefore, within the P$_1$--approximation for elastic scattering, the {\em exact} solution of
Equation~(\ref{eyieldunivariate}) is:
\begin{equation}
\label{etouret}
 \frac{\lambda_i  L}{\lambda_i+L}W_b(T)=y_L(T)-
 \frac{\lambda_iL}{\lambda_i+L} W_b(T')\otimes y_L(T-T'),
\end{equation}
where Equation~(\ref{enormbulk}) has been used.
This is the    deconvolution formula of Tougaard and Chorkendorff for which a simple recursive scheme was proposed by these authors \cite{toupr35}.

%The above derivation makes it clear that this method is based on revversion of a univariate power series (in Fourier space), which does not 
%provide a unique solution when applied to spectra represented by a bivariate power series (Eqn.(~\ref{eyieldloss}))
\subsection{II: exact reversion of a univariate power series (Ref.~\cite{vicanek} and \cite{werpia})}

Although Equation~(\ref{etouret}) is the exact solution of Equation~(\ref{eyieldunivariate}) in the
P$_1$--approximation, its validity has been questioned by several authors \cite{werpia,vicanek}, since the
P$_1$--approximation does not provide an accurate model for the {\em simulation} of REELS spectra. This becomes evident
on comparing the partial intensities in the P$_1$--approximation, with more realistic results based on the Mott cross section for elastic scattering.
Results of such calculations are displayed in Figure~\ref{fcnecs} for electrons travelling in Fe for several energies,
along with the corresponding Mott cross section \cite{yatcpc}. It is seen that the partial
intensities in the P$_1$--approximation are qualitatively different from the more realistic results based on the Mott cross section. 
This is caused by the influence of the shape of the cross section on the sequence of partial intensities
\cite{werhayreels,werepesjes}.
Realizing that the partial intensities represent the number of electrons arriving in the detector after a
given number of inelastic processes, it is clear that  the P$_1$-approximation is not a
good approach for quantitative simulation of REELS spectra (see Figure~\ref{fretsim}a below).

The exact solution of Equation~(\ref{eyieldunivariate}) for an arbitrary sequence of partial intensities, not
necessarily of the form $\alpha_n=\kappa^n$, was found independently by Vicanek \cite{vicanek} and the present author
\cite{werpia}:
\begin{equation}
\label{esoluniexact}
w_b(T)=\sum\limits_{q=1}^\infty u_q y_L^{(q)}(T)
\end{equation}
The coefficients $u_q$ in this solution are functions of the partial intensities, that are derived in Appendix~I.
This procedure was succesfully applied to experimental spectra given by a univariate power series in Fourier space, obtained by eliminating multiple 
volume scattering from them prior to application of Eqn.~(\ref{esoluniexact}) \cite{wernrelprl,wernrelsia}.

For partial intensities in the P$_{1}$--approximation ($\alpha_{n_{b}}=\kappa^{n_{b}}$), the coefficients $u_{q}$ assume a particularly simple form:
\begin{equation}
\label{euqp1}
u_{q}=\frac{(-1)^{q+1}}{\kappa}
\end{equation}
Therefore, a non-recursive alternative representation of Eqn.~(\ref{etouret}) is given by:
\begin{equation}
\label{etouretalt}
\kappa w_b(T)=\sum\limits_{q=1}^\infty (-1)^{q+1} y_L^{(q)}(T),
\end{equation}
which is sometimes useful for a theoretical analysis, although Eqn.~(\ref{etouret}) is simpler for practical applications.

\subsection{III: exact reversion of a bivariate power series (Ref.~\cite{werbivar})}
The above deconvolution procedures proceed via reversion of a univariate power series in Fourier space. However, when more
than one type of scattering contributes to the spectrum, the spectrum is represented by a bivariate power series in
Fourier space and, in consequence, a univariate power series reversion does not give the unique solution for the
involved mean free paths. This problem can be resolved \cite{werbivar} by simultaneously deconvolving a set of two
spectra taken at different energies or scattering geometries, leading to a different sequence  of partial
intensities. In doing so, it is assumed that the {\em normalized}Êdifferential mean free path for volume and
surface  scattering is independent of the energy and scattering geometry, i.e. that the normalized mean free paths in the
two spectra  are identical [see Eqn.~(\ref{enormdiimfpdsep})]. This turns out to be a very reasonable approximation \cite{werbivar,werqsasia}.
Then, the unique solution for the mean free path for surface and bulk scattering is given by the expression:
\begin{eqnarray}
\label{ereversionrealspace}
{w}_b (T)
&=&
\sum\limits_{p=0}^\infty
\sum\limits_{q=0}^\infty
u_{p,q}^b 
{y}_{L,1}^{(p)} (T')\otimes
{y}_{L,2}^{(q)}(T-T')
\nonumber\\
{w}_s(T)
&=&
\sum\limits_{p=0}^\infty
\sum\limits_{q=0}^\infty
u_{p,q}^s 
{y}_{L,1}^{(p)}(T')\otimes
{y}_{L,2}^{(q)}(T-T')
\end{eqnarray}
where the coefficients $u_{pq}^b$ and $u_{pq}^s$ are again functions of the partial intensities. The algorithm to obtain
these coefficients is given in Appendix II.

%%%%%%%%%%%%%%%%%%%%%%%%%%%%%%%%%%%%%%%%%%%%%%%%%%%%%%%%%%%%%%%%%%%%%%
\section{Results and Discussion}
In order to compare the procedures described above, they  were first applied to results of model calculations (as described in Ref.~\cite{werbivar}) and
subsequently  experimental data were deconvoluted with the method of Tougaard and Chorkendorff (TC) and with the reversion
of a bivariate power series.

The procedure used to acquire the experimental data  has been described in detail before \cite{werepesjes}. REELS data
for Fe, Pd and Pt were measured in the  energy range between 300 and 3400~eV for normal incidence and an off-normal
emission angle of 60$^\circ$, using a hemispherical analyzer operated in the constant analyzer energy mode giving a width
of the elastic peak of 0.7~eV. Count  rates in the elastic peaks were kept well below the saturation count rate of the
channeltrons  and a dead time correction was applied to the data. For each material the optimum energy combination for
the retrieval procedure of  two loss spectra was determined by inspection of the partial intensities. For most materials
the  optimum energy combination was (1000-3000~eV) \cite{werbivar}.

\begin{figure}[htb]
%\begin{center} 
\ifpdf
{\includegraphics[width=17.0cm]{fcnecs.pdf}}
\else
\epsfxsize = 0.8\hsize
\epsffile{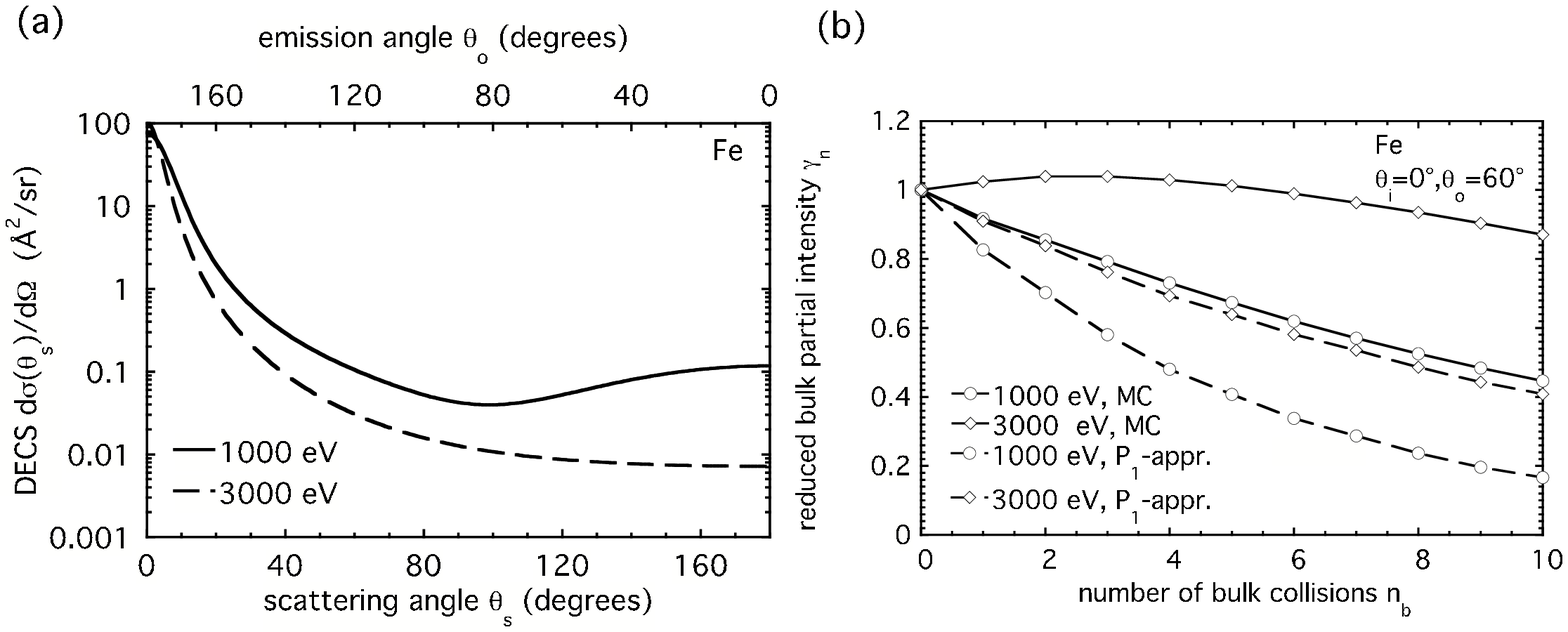}
\fi
\caption{%
(a) Differential Mott cross section for elastic scattering for Fe for two electron energies. (b) Reduced partial
intensities $\alpha_{n_b}$ for volume scattering for several energies in Fe. Solid curves: Monte
Carlo calculations based on the Mott scattering cross section. Dashed curves: P$_1$--approximation (Equation~(\ref{epip1}))
}
\label{fcnecs}
%\end{center}
\end{figure}

All simulated data were calculated for the scattering geometry used in the experiment with the method described in Refs.~\cite{werbivar,werepesjes}.
In a first step, simulated spectra were generated neglecting surface excitations and using the bulk partial intensities
calculated within the $P_{1}$--approximation. In this case  method I exactly returned the model  DIIMFP. When the more
realistic partial intensities based on the Mott cross section are used (see Figure~\ref{fcnecs}), the simulated spectra
are significantly different.
 However, the results of method I and II for the
shape of the retrieved DIIMFP was indistinguishably similar, the only difference is that the absolute value of the
DIIMFP differs  since the first and second order partial intensities are different. The
reason why both methods give essentially the same results for the shape of the DIIMFP is easy to understand on inspection
of the partial intensities in Figure~\ref{fcnecs}. Although the sequence of partial intensities obtained from MC
calculations is qualitatively different from the one obtained in the P$_{1}$--approximation, the MC-partial
intensities for the two energies shown are also to a good approximation of the form $\alpha_{n}=\kappa^{n}$ (with a different value for $\kappa$), at least
for the first few scattering orders $n\sim1-5$. Therefore, method I essentially results in the correct deconvolution
for any spectrum that can be represented as a univariate power series of the DIIMFP in Fourier space. This is true for
any realistic sequence of partial intensities for the first few scattering orders, except for some really pathological
cases \cite{werpia}.

\begin{figure}[htb]
%\begin{center} 
\ifpdf
{\includegraphics[width=17.0cm]{fretfin.pdf}}
\else
\epsfxsize = 0.8\hsize
\epsffile{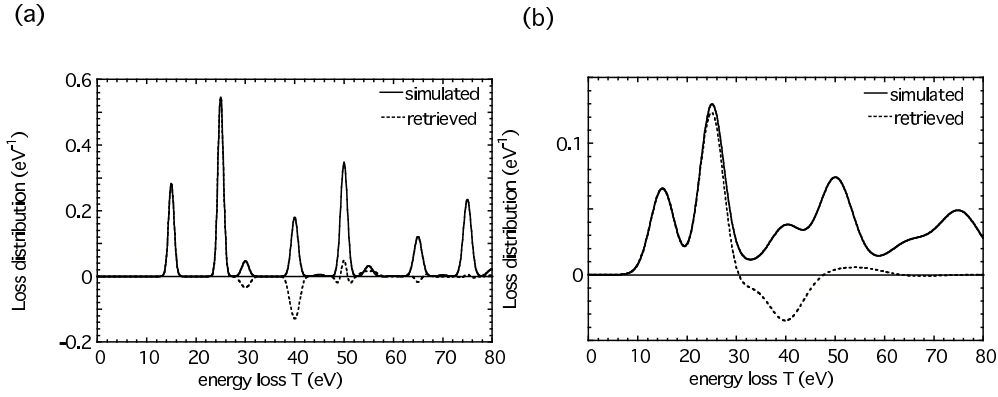}
\fi
\caption{%
Simulated spectra (solid lines) for a model DIIMFP and DSEP represented by a Gaussian peak shape with a mean energy loss of 25 and 15~eV, respectively. The dashed curves 
are the result of application of Tougaard and Chorkendorff's algorithm Eqn.~(\ref{etouret}). It is seen that the latter does not eliminate the higher order terms. (a.) a model plasmon width of 0.4~eV was used. (b)
 a model plasmon width of 2.4~eV was used.
}
\label{fretfin}
%\end{center}
\end{figure}

However, when method I is applied to simulated data in which surface excitations are not ignored, the results are no
longer interpretable in a straightforward way. This can be seen in Figure~\ref{fretfin} that shows model calculations
for loss spectra with plasmon peaks modelled by Gaussian peak shapes (15~eV for the surface-- and 25~eV for the bulk plasmon).
The width of the plasmon loss peaks was taken to be 0.4~eV in (a) and 2.4~eV in (b). The retrieved loss distributions clearly contain significant intensity 
from higher order scattering.

This becomes understandable by analyzing the  loss distribution retrieved by method I. In order to do so, the spectrum [Eqn.~(\ref{eyieldloss})] is inserted into 
Eqn.~(\ref{etouretalt}), giving:
\begin{equation}
\label{etourec}
\kappa w^{TC}(T)=\alpha_{1,0}w_{b}(T)
+\alpha_{0,1}w_{s}(T)
-\alpha_{1,1}w_{b}(T')\otimes w_{s}(T-T')-\alpha_{2,0}w_{s}^{(2)}(T)+{\cal O}(w^{(3)})
\end{equation}
where Eqns.~(\ref{epiuncorr}) and (\ref{episurf}) were used.
Thus, it is found that the loss distribution of Tougaard and Chorkendorff contains a negative contribution equal to the second order mixed and surface scattering term.
This is clearly observed in Figure~\ref{fretfin}a. It is also seen that the higher orders decrease rapidly with the scattering order. 
 Since the average number of surface excitations is usually quite small compared to unity, the
contribution of surface excitations  may become negligible for high enough energies, but the essential point is that the second
order mixed term is always appreciable in magnitude since it remains comparable to the single scattering partial
intensities for surface (and bulk) scattering. This finding also explains the spurious negative excursions for energies corresponding to one surface and one bulk loss 
in the retrieved loss distributions  in Ref.~\cite{toupr35} and similar works published in the past twenty years.

\begin{figure}[htb]
%\begin{center} 
\ifpdf
{\includegraphics[width=17.0cm]{fretsim.pdf}}
\else
\epsfxsize = 0.8\hsize
\epsffile{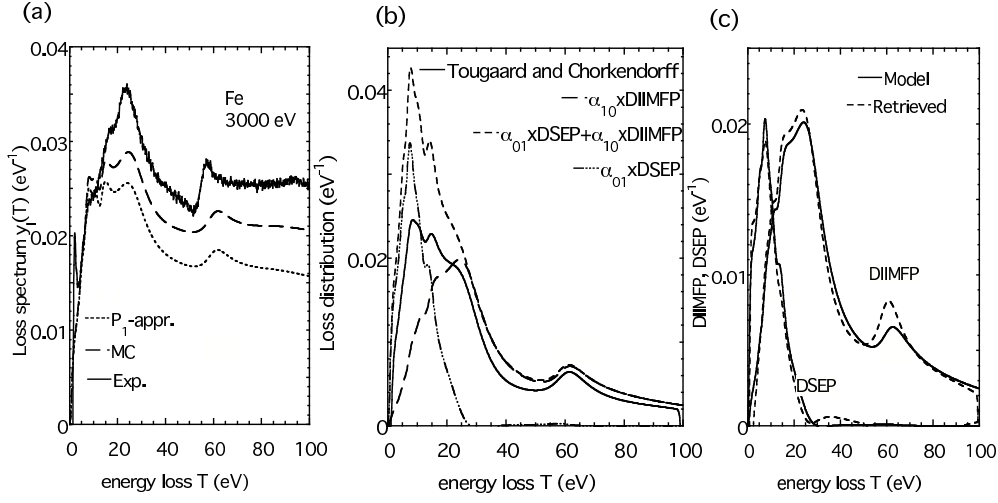}
\fi
\caption{%
(a) experimental REELS spectrum for 3000~eV-electrons backscattered from an Fe-surface (solid curve),
spectrum simulated with Equation~(\ref{eyieldloss}) based on the Mott cross section for elastic scattering (dashed curve)
and simulated spectrum within the P$_1$--approximation for elastic scattering (dotted curve). (b) loss distribution retrieved from
the simulated spectrum in (a) based on the Mott cross section for elastic scattering using the Method of Tougaard and
Chorkendorff \cite{toupr35} (solid curve). The long-dashed curve represents the single volume scattering
contribution of the simulated spectrum, the dash-dotted curve is the contribution of single surface scattering. The 
short-dashed curve is the total single scattering contribution in the spectrum in (a). (c.) normalized DIIMFP and DSEP retrieved
from the simulated spectra in Figure~\ref{floss}(a) using the bivariate power series reversion \cite{werbivar} (dashed
curves) compared with the DIIMFP and DSEP used in the simulation (solid curves). 
 }
\label{fretsim}
%\end{center}
\end{figure}

Similar results were obtained for more realistic simulations for 500~eV electrons in Fe, shown in Figure~\ref{fretsim}.
In Figure~\ref{fretsim}a the dotted curve represents the simulation in
the P$_{1}$--approximation, while the dashed curve is based on Monte Carlo (MC) calculations for the pathlength
distribution from which the partial intensities are derived via Equation~(\ref{epipld}). The latter are seen to agree
better with the experimental data (represented by the noisy solid curve). The reason is that the realistic Mott cross section for elastic scattering 
can not be described by a first order Legendre polynomial, as implicitly assumed in the P$_{1}$--approximation (see  Figure~\ref{fcnecs}a).
In  Figure~\ref{fretsim}b, the 
distribution retrieved with method I is shown as the solid curve and is compared with
the single surface and volume scattering contribution to the spectrum, represented by the chain--dashed and
long-dashed curves, respectively. The short dashed curve is the total single scattering distribution of the model
spectrum, i.e. $\alpha_{1,0}w_{b}(T)+\alpha_{0,1}w_{s}(T)$. It is seen that the retrieved loss distribution is lower
than the single scattering loss distribution and that it can {\em not } be represented by  any linear combination of
the single scattering contributions.  Thus, the method of
Tougaard and Chorkendorff never gives a {\em single}Êscattering loss distribution
when surface scattering is not negligible. 

Finally, the method suggested by Chen \cite{chenprb58} deserves to be addressed in this context. This author proposed
to extract the DIIMFP and the DSEP by using the "effective" cross section of Tougaard and Chorkendorff obtained from
REELS spectra taken at two different energies. This method is thus experimentally similar to the bivariate power
series reversion, however, the results are quite different, as will be shown below.  To eliminate the first order bulk
contribution, giving a retrieved distribution with the first order DSEP as the leading term,  two Tougaard-Chorkendorff
cross sections $w^{TC}_{1}$ and $w_{2}^{TC}$  are subtracted from each other after one of them is multiplied with
$\nu=\alpha_{1,0}/\beta_{1,0}$.  Here $\alpha_{n_{b},n_{s}}$ are the partial intensities of spectrum 1 and
$\beta_{n_{b},n_{s}}$ are the partial intensities of spectrum 2. This gives:
\begin{equation}
\label{echenrec}
\kappa_{1} w^{TC}_{1}(T)-\kappa_{2} w^{TC}_{2}(T)=\Delta_{1,0}w_{b}(T)+\Delta_{0,1}w_{s}(T)-\Delta_{1,1}w_{b}(T')\otimes w_{s}(T-T')-\Delta_{2,0}w_{s}^{(2)}(T)+{\cal O}(w^{3})
\end{equation}
where $\Delta_{n_{b},n_{s}}=\alpha_{n_{b},n_{s}}-\nu\beta_{n_{b},n_{s}}$.
It is seen, also in this case, that the second order mixed term is always  of the same order of  
magnitude as the first order surface term in the resulting (difference) spectrum:
\begin{equation}
-\frac{\Delta_{1,1}}{\Delta_{0,1}}\simeq -1,
\end{equation}
So that in this method the mixed terms also make it impossible to recover the single scattering loss distributions from
two "effective" cross sections obtained from REELS spectra taken at two energies (or scattering geometries). The analysis
of Chen \cite{chenprb58} is quite different from the present one. In the work by this author, the mixed terms are
"eliminated" by replacing the Fourier transform of the bulk DIIMFP ($\widetilde{\mu}_{B}^{0}(s)$ in the notation of that
paper) in Eq.(45) by the same quantity in real space in Eqns.~(46)-(51). In this way this procedure  ignores
the mixed terms, but unfortunately, they are of the same order of magnitude as the first order surface and bulk term.

The result of the  retrieval using the bivariate power series reversion is compared in Figure~\ref{fretsim}c (dashed
lines) with the model DIIMFP and DSEP for 3000 eV (solid lines). In this case the retrieved
distributions for surface and
bulk scattering are separated and   are very similar to the model distributions. The differences that can be observed
are due to the assumption that the { normalized} DIIMFP and DSEP are independent of the energy of the probing
electron and the direction of surface crossing. Although this is a good approximation, it is not exactly true and
hence the retrieved distributions agree with the model loss distribution (for one energy) within the accuracy of this
approximation. If the model DIIMFP and DSEP are taken to be the same for both energies and for the incoming and
outgoing direction of surface crossing, the retrieved distributions {\em exactly} agree with the model distributions,
since Eqn.~(\ref{ereversionrealspace}) is the {\em exact} solution of Eqn.~(\ref{eyieldloss}).

\begin{figure}[htb]
%\begin{center} 
\ifpdf
{\includegraphics[width=17.0cm]{fxretsim.pdf}}
\else
\epsfxsize = 0.8\hsize
\epsffile{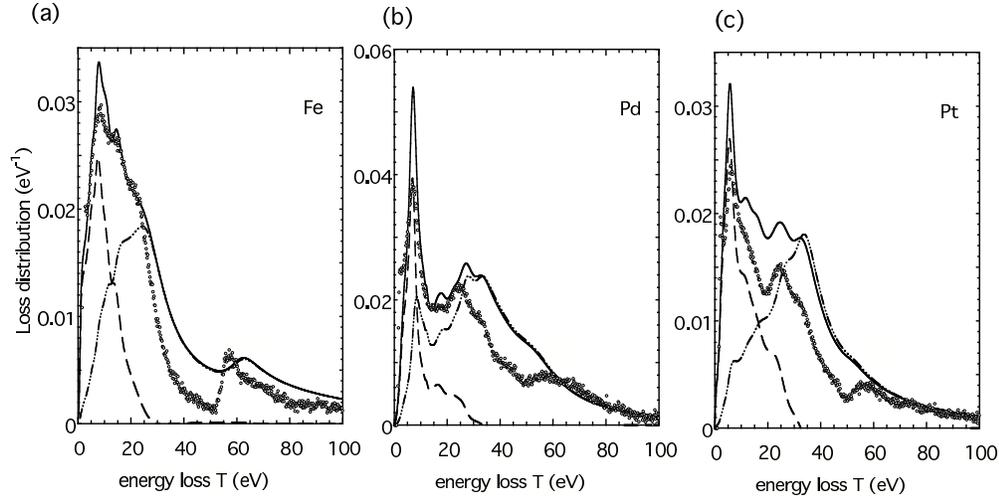}
\fi
\caption{%
Loss distribution obtained with the method of Tougaard and Chorkendorff \cite{toupr35}(open circles). The dash-dotted
and dashed  curves are the normalized DIIMFP  and DSEP calculated with the formulae in \cite{tungpr49} and multipled with the
according single scattering partial intensities,
the solid curve represents their sum. In all cases, it can be seen that the retrieved spectrum is {\em not}Êa linear
combination of the single scattering terms. (a) Fe; (b) Pd; (c) Pt.  }
\label{fxretsim}
%\end{center}
\end{figure}

Application of Tougaard and Chorkendorff«s method to the 1~keV experimental  spectra of Fe, Pd and Pt is shown in
Figure~\ref{fxretsim} as open circles. The dashed and dash-dotted curves represent the theoretically expected
contribution of surface and volume scattering, the solid curve is their sum. 
%The theoretical single scattering 
%loss distributions shown here are found to agree quantitatively with the results of the method of bivariate power
%series reversion (see Figure~\ref{floss} below).
For all three materials, the
experimentally retrieved loss distribution in Figure~\ref{fxretsim} is  seen to be significantly lower than the total
single scattering contribution (solid curves) for energies $\alt 80$~eV while for higher energy losses  they become
similar to the theoretically expected result.
 This makes it clear that the loss distributions retrieved by the method of Tougaard and Chorkendorff  are not a linear
combination of the single scattering surface and volume loss
distributions being attributable (mainly) to the negative second order mixed term, as discussed above. Thus, although
the retrieval of these data from experimental results is very simple, the usefullness of this procedure is
questionable since quantitative interpretation of the results is by no means straightforward.

\begin{figure}[htb]
%\begin{center} 
\ifpdf
{\includegraphics[width=17.0cm]{floss.pdf}}
\else
\epsfxsize = 0.8\hsize
\epsffile{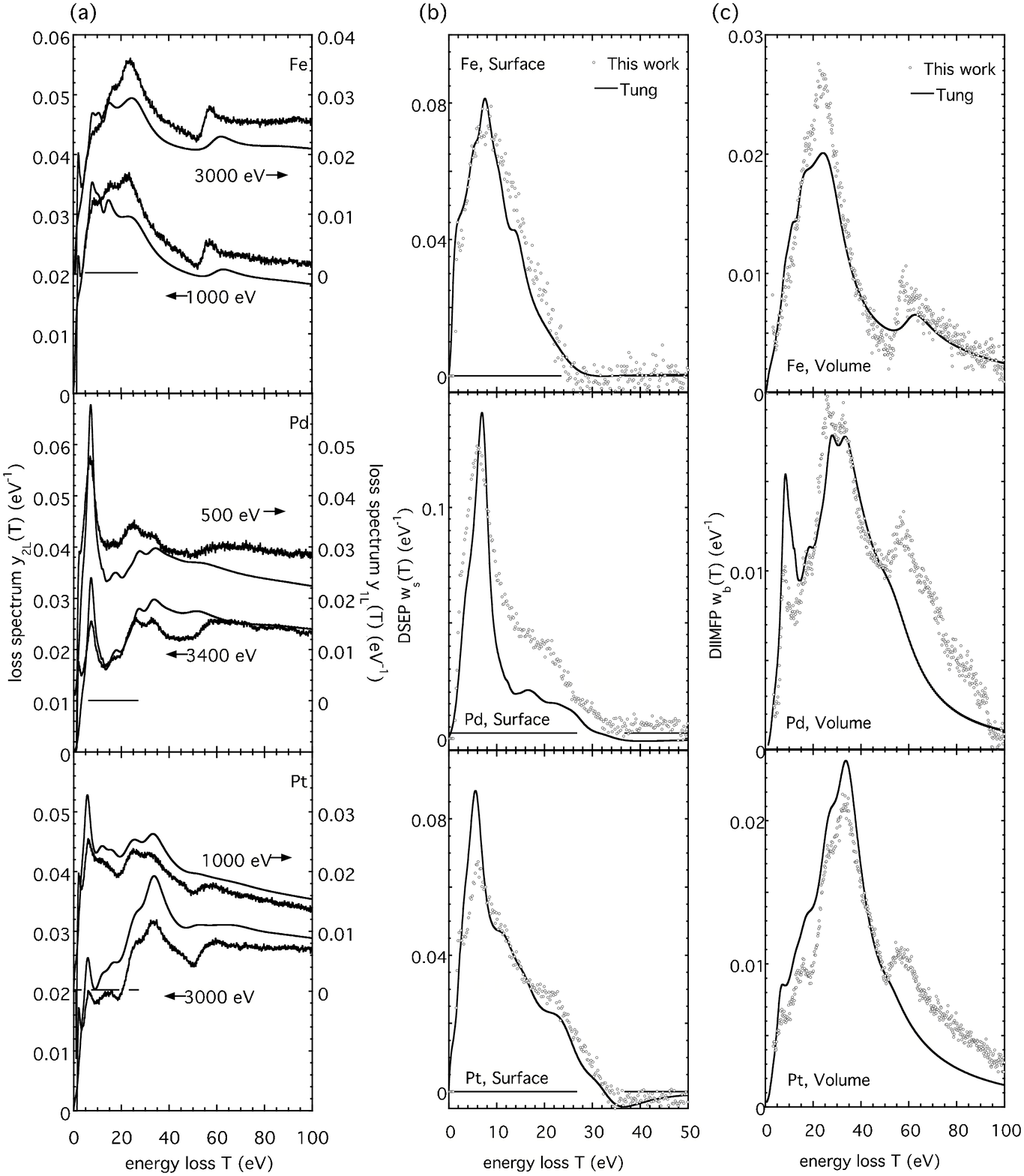}
\fi
\caption{%
(a) Measured energy loss spectra  for Fe, Pd and Pt for the energies indicated in the legend after removal of the elastic peak  (noisy curves). The
smooth curves are the simulated loss spectra using theoretical shapes for the DSEP and DIIMFP. 
(b) Retrieved normalized DSEP (open circles). 
 (c) Retrieved normalized DIIMFP (open
circles). The solid curves in (b) and (c) represent the result based on the theory  by Tung and coworkers
\cite{tungpr49}.
 }
\label{floss}
%\end{center}
\end{figure}

The method of reversion of a bivariate power series in Fourier space is illustrated in Figure~\ref{floss}
for Fe (1000-3000~eV), Pd (500-3400~eV) and Pt (1000-3000~eV). Figure~\ref{floss}a compares the measured loss spectra
(i.e. the spectra after division by the area of the elastic peak and after removal of the latter) as noisy solid
curves with simulated spectra (smooth solid curves).  Reasonable agreement between simulation and experiment
is seen in all cases, but significant deviations can  also be observed. For example, the ionization edges at
$\sim$60~eV are not exactly reproduced by the simulation in all cases. This is attributable to the fact that it is
difficult to fit ionization edges with a Drude-Lindhard type expansion of the dielectric function \cite{tungpr49}.

In Figure~\ref{floss}b and c the DSEP  and DIIMFP  retrieved with the bivariate power series reversion are represented
by the open circles and compared with result based on the theory of Tung and coworkers \cite{tungpr49} (solid curves). 
In all cases the agreement is quite good, except in the
vicinity of ionization edges. Furthemore, the main feature in the DIIMFP of Fe is a bit sharper than  predicted by
theory and the DSEP for Pd is significantly higher in the energy range between 10--30~eV. Note that the retrieved
distributions were not scaled in any way, the normalized loss distributions are compared on an absolute scale.

The DSEP always exhibits a negative excursion for an energy corresponding approximately to the average volume
loss. This feature, that has a clear physical explanation \cite{wernrelprl},  is also seen in the theoretical results and
is a consequence of the coupling of surface and bulk excitations that are essentially different (orthogonal) modes of the
inelastic interaction. The coupling term is included in the DSEP, as defined in the present work \cite{werzemsep,werbivar},
and it is always negative. So the DSEP is in fact a difference of two probabilities, the pure surface excitation term and
the coupling term that takes into account the orthogonality of the surface and bulk modes, and the resulting DSEP
exhibits a negative excursion whenever the coupling term exceeds the pure surface term.

\begin{figure}[htb]
%\begin{center} 
\ifpdf
{\includegraphics[width=17.0cm]{fcmppdpt.pdf}}
\else
\epsfxsize = 0.8\hsize
\epsffile{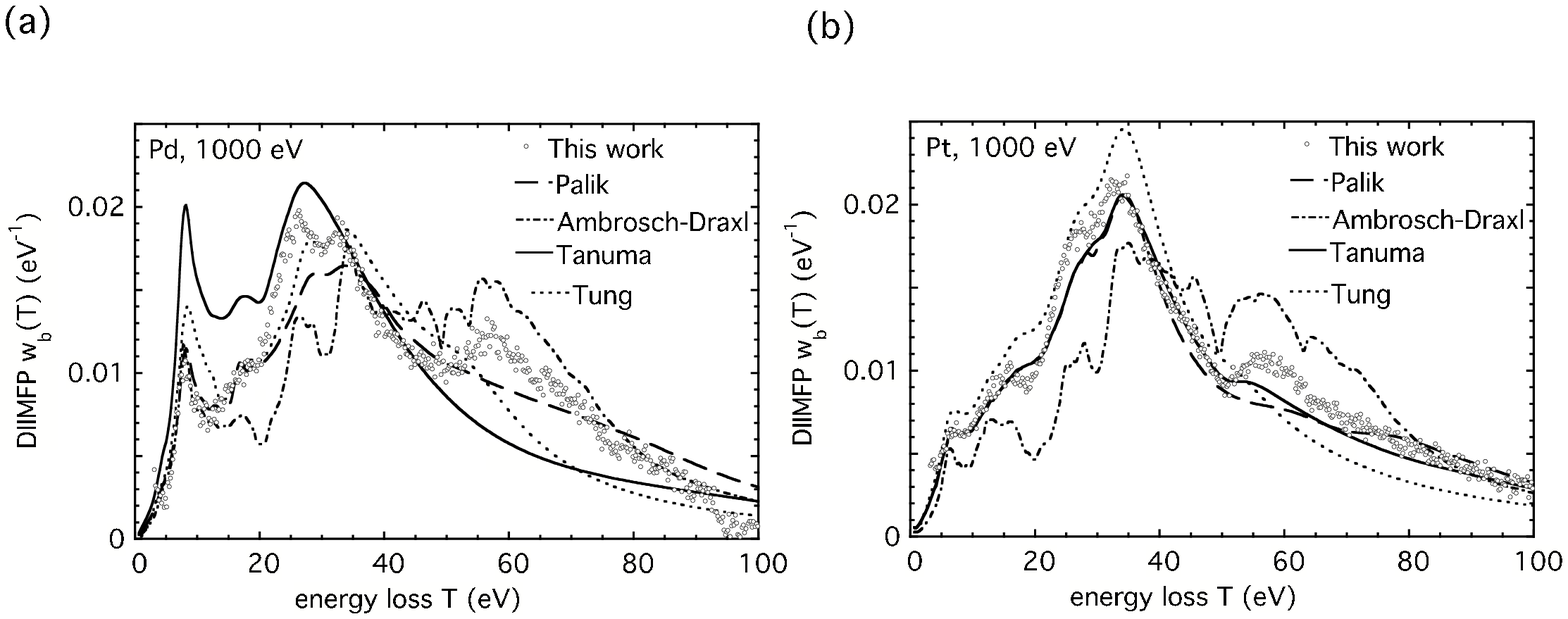}
\fi
\caption{%
Comparison of the DIIMFP obtained in the present work (open circles)  with theoretical calculations
using optical data taken from different sources \cite{tungpr49,palik,ambroschpr,tanumaoptdat} as indicated in the
legend.
(a) Pd, 1000 eV. (b) Pt, 1000 eV. 
}
\label{fcmp}
%\end{center}
\end{figure}

Finally, in Figure~\ref{fcmp} the present results for the volume  single scattering loss distribution is compared with
results based on optical data from various sources in the literature \cite{tungpr49,palik,ambroschpr,tanumaoptdat}. 
The literature data exhibit significant discrepancies.  For Pd, the present results agree well with those of Tung
\cite{tungpr49} for energies below $\sim$50~eV. For Pt, the present results excellently match the data by Tanuma
\cite{tanumaoptdat} and Palik \cite{palik}. In both cases, the ionization edge is more pronounced in the present
data. This feature is well reproduced by first principles calculations based on density functional theory calculations beyond the ground state for both materials
(dash-dotted lines) \cite{ambroschpr}.

\section{Summary and conclusions} 
The single scattering loss distribution for inelastic electron scattering was extracted from experimental data for Fe,
Pd and Pt. This was done by means of a recently proposed procedure \cite{werbivar} in which it is assumed that the
{\em normalized} differential mean free path for inelastic scattering (DIIMFP) does not depend on the energy of the
probing electron. Furthermore, it is assumed that the {\em normalized} differential surface excitation probability
(DSEP) does not depend on the surface crossing direction and the primary energy. Under these assumptions, a set of two
spectra with a different sequence of partial intensities can be simultaneously deconvoluted to give the normalized
true single scattering loss distributions. The deconvolution procedure proceeds via the reversion of a bivariate 
power series in Fourier space. Experimentally, two spectra with
a different sequence of partial intensities can be obtained by measuring the loss spectrum at two different energies,
as studied in the present work, or, alternatively, for two different scattering geometries. The latter case, which was
not studied here, is a bit more demanding experimentally, but has the great advantage that the DIIMFP and DSEP of the
two spectra are indeed identical. In other words, it is then only the angular dependence of the DSEP  (for the in- and outgoing beam)
that may slightly distort the deconvolution results. However, the present results based on spectra taken at two
different energies also compare very well with theoretical results since the energy dependence of the normalized
single scattering loss distributions is quite weak in reality.

A detailed comparison was made with earlier methods based on reversion of a univariate power series. The most well
known and frequently used deconvolution procedure by Tougaard and Chorkendorff \cite{toupr35} belongs to this class
of procedures. It was shown, both theoretically and on the basis of model calculations, that when such a method is
applied to a spectrum described by a bivariate power series (in the surface and volume single scattering loss
distribution), it results in a not very well defined mixture of multiple volume, surface and mixed scattering terms. In
particular,  the second order mixed term (corresponding to electrons that experience one surface and one
volume energy loss) gives a strong negative contribution in the resulting loss distribution. This explains the spurious
negative excursion at  energies corresponding to one bulk and one surface loss in spectra deconvoluted by the method of
Tougaard and Chorkendorff that have been published over the past twenty years (see e.g., Ref.~\cite{toupr35}).

Finally, and most importantly, the above findings confirm the theoretical model for REELS based on a
bivariate power series in the DSEP and the DIIMFP  in Fourier space \cite{werbivar} with unprecedented detail.

\section{Acknowledgments}
The author is grateful to Dr. S. Tanuma (Ref.~\cite{tanumaoptdat}) and Dr. C. Ambrosch-Draxl (Ref.~\cite{ambroschpr})
for making their optical data  available before publication.  Financial support of the present work
by the Austrian Science Foundation FWF  through Project No. P15938-N02 is gratefully acknowledged

\section{Appendix}
\subsection{Reversion of a univariate powers series}
We consider the loss spectrum  ${y}_L$ written as a power series of the variable ${w}$ in Fourier space (i.e. a
multiple self-convolution in real space):
\begin{equation}
\label{e_app1}
{y}_L(T)=\sum\limits_{n=0}
^\infty \alpha_n  {w}^{(n)}(T)
\end{equation}
where $\alpha_0=0$ since it is assumed that the elastic peak has been removed from the spectrum.
Since the leading term in ${y}_L$ is the first power of ${w}$, the solution must be of the form:
\begin{equation}
\label{solution}
{w}(T)=\sum\limits_{q=1}^\infty u_q {y}_L(T)^{(q)}.
\end{equation}
Inserting  Eq.\ref{e_app1} into this ansatz one finds:
\begin{equation}
{w}(T)=\sum\limits_{q=1}^\infty u_q 
\Big\{ 
\sum\limits_{n=1}^\infty \alpha_n {w}^{(n)}(T)\Big\}^{(q)},
\end{equation}
and, equating coefficients of equal powers in the above equation:
\begin{eqnarray}
\label{solution1}
1&=&u_1 \alpha_1\nonumber\\
0&=&u_1\alpha_2+u_2\alpha_1\alpha_1\nonumber\\
0&=&u_1\alpha_3+u_2(\alpha_1\alpha_2+\alpha_2\alpha_1)+u_3\alpha_1\alpha_1\alpha_1\nonumber\\
0&=&u_1\alpha_4+u_2(\alpha_2\alpha_2+\alpha_1\alpha_3+\alpha_3\alpha_1)+u_3(\alpha_2\alpha_1
\alpha_1+\alpha_1\alpha_2\alpha_1+\alpha_1\alpha_1\alpha_2)+u_4 \alpha_1\alpha_1\alpha_1\alpha_1\nonumber\\
...\nonumber
\end{eqnarray}
the solution for the unknown coefficients $u_q$ is easily seen and  can be found in any standard textbook
\cite{abramowitz,bronstein} in a slightly different and more  explicit albeit less general form.
 The interesting point in the present notation, however, is that
the subscripts in the coefficients of $u_q$ in the above equations represent the sum of all permutations of the 
partitions of $n$ (the order) with $q$ factors in them. 
\subsection{Reversion of a bivariate power series.}
When {\em two } loss spectra with a 
different sequence of partial intensities are measured,
\begin{eqnarray}
\label{etwospectra}
{y}_{L,1}(T)
&=&
\sum\limits_{n_s=0}^\infty
\sum\limits_{n_b=0}^\infty
\alpha_{n_s , n_b} 
{w}_b^{(n_b)}(T')\otimes
{w}_s^{(n_s)}(T-T')
\nonumber\\
{y}_{L,2}(T)
&=&
\sum\limits_{n_s=0}^\infty
\sum\limits_{n_b=0}^\infty
\beta_{n_s , n_b} 
{w}_b^{(n_b)}(T')\otimes
{w}_s^{(n_s)}(T-T'),
\end{eqnarray}
with $\alpha_{0,0}=\beta_{0,0}=0$, a unique solution for the unknowns $w_s(T)$ and $w_b(T)$ can be established by 
reversion of this bivariate power series (in Fourier space). 

Formally, the reversion of this bivariate power series is effected by the  expansion:
\begin{eqnarray}
\label{eformalreversion}
{w}_b(T)
&=&
\sum\limits_{p=0}^\infty
\sum\limits_{q=0}^\infty
u_{p,q}^b 
{y}_{L,1}^{(p)}(T')\otimes
{y}_{L,2}^{(q)}(T-T')
\nonumber\\
{w}_s(T)
&=&
\sum\limits_{p=0}^\infty
\sum\limits_{q=0}^\infty
u_{p,q}^s 
{y}_{L,1}^{(p)}(T')\otimes
{y}_{L,2}^{(q)}(T-T')
\end{eqnarray}
with $u_{0,0}^b=u^s_{0,0}=0$.
This can be seen by substituting  Equation~(\ref{etwospectra}) back into  Equation~(\ref{eformalreversion}) and equating
coefficients of equal powers of ${w}_b$ and ${w}_s$. This  gives the equations for the unknown
coefficients $u_{p,q}^b$ and $u^s_{p,q}$. 

In doing so, one is faced with the problem of evaluating the convolution of 
${y}_{L,1}^{(p)}(T)$ with
${y}_{L,2}^{(q)}(T)$ that can be expressed as:
\begin{equation}
\label{etensorgamma}
{y}_{L,1}^{(p)}(T')\otimes
{y}_{L,2}^{(q)}(T-T')=
\sum\limits_{n_s=0}^\infty
\sum\limits_{n_b=0}^\infty
\gamma_{p,q,n_b,n_s}
{w}_b^{(n_b)}(T')\otimes
{w}_s^{(n_s)}(T-T')
\end{equation}
The coefficients $\gamma_{p,q,n_b,n_s}$  are found to be  given by: 
\begin{equation}
\label{egammacomponents}
\gamma_{p,q,n_b,n_s}=
(\alpha_p,\beta_q)_{(n_b,n_s)}^{(p+q)}
\end{equation}
These coefficients are equal to the sum of all possible terms with $p$ factors in $\alpha_{k,l}$ and  $q$ factors in
$\beta_{m,n}$, whose indices "add up" to the target index combination $(n_s,n_b)=(k+m,l+n)$. For example:
\begin{eqnarray}
\label{egammaexamples}
(\alpha_2)_{(2,1)}^{(2)}&=&
\alpha_{1,0}\alpha_{1,1}+
\alpha_{0.1}\alpha_{2,0}+
\alpha_{1,1}\alpha_{1,0}+
\alpha_{2,0}\alpha_{0,1}
\nonumber\\
(\alpha_1,\beta_2)_{(1,2)}^{(3))}&=&
\alpha_{0,1}\beta_{0,1}\beta_{1,0}+
\alpha_{0,1}\beta_{1,0}\beta_{0,1}+
\alpha_{1,0}\beta_{0,1}\beta_{0,1}
\end{eqnarray}
Obviously,  one has
\begin{equation}
\label{egammais0} 
\gamma_{p,q,n_b,n_s}=0~~~~~~~~~~~~for~all~p+q>n_s+n_b
\end{equation}
since the target index combination $(n_s,n_b)$ can only be
produced by a number of factors less than or equal to $(n_s+n_b)$ when $\alpha_{0,0}=\beta_{0,0}=0$ [see
Equation~(\ref{etwospectra})]. Furthermore, one has:
\begin{eqnarray}
\label{ethis}
\gamma_{1,0,n_b,n_s}&=&\alpha_{n_b,n_s}
\nonumber\\
\gamma_{0,1,n_b,n_s}&=&\beta_{n_b,n_s}
\end{eqnarray}
Based on these guidelines, the coefficients $\gamma_{p,q,n_b,n_s}$ can
readily be established by means of a recursive algorithm for any sequence of partial intensities $\alpha_{n_b, n_s} $ and
$\beta_{n_b , n_s} $.

Inserting Equation~(\ref{etwospectra}) into Equation~(\ref{eformalreversion}) and equating coefficients, one
finds a set of  two equations with two unknowns  for the first order bulk coefficients:
\begin{eqnarray}
\label{eupqbeq}
1&=&
\sum\limits_{p=0}^\infty
\sum\limits_{q=0}^\infty
u^b_{p,q}\gamma_{p,q,1,0}
%=u_{0,1}\gamma_{0,1,1,0}+u_{1,0}\gamma_{1,0,1,0}
=u^{b}_{0,1}\beta_{1,0}+u^{b}_{1,0}\alpha_{1,0}
\nonumber\\
0&=&
\sum\limits_{p=0}^\infty
\sum\limits_{q=0}^\infty
u^b_{p,q}\gamma_{p,q,0,1}=
%u_{0,1}\gamma_{0,1,0,1}+u_{1,0}\gamma_{1,0,0,1}=
u^{b}_{0,1}\beta_{0,1}+u^{b}_{1,0}\alpha_{0,1}
\end{eqnarray}
The solution is:
\begin{eqnarray}
\label{eupqb1}
u_{1,0}^b&=&\frac{\beta_{0,1}}{\beta_{0,1}\alpha_{1,0}-\beta_{1,0}\alpha_{0,1}}
\nonumber\\
u_{0,1}^b&=&\frac{\alpha_{0,1}}{\alpha_{0,1}\beta_{1,0}-\alpha_{1,0}\beta_{0,1}}
\end{eqnarray}
Similarly, for the first order surface coefficients one finds:
\begin{eqnarray}
\label{eupqs1}
u_{1,0}^s&=&\frac{\beta_{1,0}}{\beta_{1,0}\alpha_{0,1}-\beta_{0,1}\alpha_{1,0}}
\nonumber\\
u_{0,1}^s&=&\frac{\alpha_{1,0}}{\alpha_{1,0}\beta_{0,1}-\alpha_{0,1}\beta_{1,0}}
\end{eqnarray}
The equation determining the higher order ($p+q>1)$ coefficients
\begin{equation}
\label{eupqhigh1}
0=
\sum\limits_{p=0}^\infty
\sum\limits_{q=0}^\infty
u_{p,q}
\gamma_{p,q,n_b,n_s}
\end{equation}
can be split into two parts by using property (\ref{egammais0}),  changing the order of the summation and writing:
\begin{equation}
\label{eupqhigh2}
0=
\sum\limits_{p=0}^{n_s+n_b-1}
\sum\limits_{q=0}^p
u_{q,p-q}
\gamma_{q,p-q,n_b, n_s}
+
\sum\limits_{p=0}^{n_s+n_b}
u_{p,n_s+n_b-p}
\gamma_{p,n_b+n_s-p,n_b,n_s}
\end{equation}
The first term represents a $(n_s+n_b+1)$--dimensional vector containing the coefficients $u_{p<n_b, q<n_s}$, which
have been established during the previous step of the algorithm.
The latter term is the product of a square matrix with the same dimension and the unknown vector $(u_{0,n_b+n_s},
u_{1,n_b+n_s-1},  u_{2,n_b+n_s-2}, ...u_{n_b+n_s,0})$. Thus, for each value of $n_b+n_s$, the corresponding coefficients
are obtained by solution of a system of $n_b+n_s+1$ linear equations with as many unknowns. Consecutively performing this procedure for
values of the total scattering order $n_b+n_s=1,2,....n_{max}$, where $n_{max}$ is the collision order where convergence
of the series Equation~(\ref{eformalreversion})  is attained, leads to the desired coefficient matrices $u_{p,q}^b$ and
$u_{p,q}^s$. Note that Eqns.~(\ref{eupqhigh1}) and (\ref{eupqhigh2}) are identical for the surface and bulk coefficients,
only the calculation for the first order term is different [see Eqns.~(\ref{eupqb1}) and (\ref{eupqs1})].
Although faster algorithms for reversion of multivariate
power series are available \cite{raney,brent,cheng}, these are quite complex and were not considered here since
convergence of the proposed algorithm is typically attained for $n_{max}\le7$ for energy loss ranges extending up to
100~eV.

%\bibliography{/Volumes/WolfTrance/TranceVair/tex/Allref}

\newpage
%%%%%%%%%%%%%%%%%%%%%%%%%%%%%%%%%%%%%%%%%%%%%%%%%%%%%%%%%%%%%%%%%%%%%%%%%%%%%%%%%%%%%%%%%%%%%%%%%%%%%%%%%%%%%%%%%%%%%%%%%%%%%%%%%%%%%%%%%%%%%%%%%
%%%%%%%%%%%%%%%%%%%%%%%%%%%%%%%%%%%%%%%%%%%%%%%%%%%%%%%%%%%%%%%%%%%%%%%%%%%%%%%%%%%%%%%%%%%%%%%%%%%%%%%%%%%%%%%%%%%%%%%%%%%%%%%%%%%%%%%%%%%%%%%%%

%%%%%%%%%%%%%%%%%%%%%%%%%%%%%%%%%%%%%%%%%%%%%%%%%%%%%%%%%%%%%%%%%%%%%%%%%%%%%%%%%%%%%%%%%%%%%%%%%%%%%%%%%%%%%%%%%%%%%%%%%%%%%%%%%%%%%%%%%%%%%%%%%
%%%%%%%%%%%%%%%%%%%%%%%%%%%%%%%%%%%%%%%%%%%%%%%%%%%%%%%%%%%%%%%%%%%%%%%%%%%%%%%%%%%%%%%%%%%%%%%%%%%%%%%%%%%%%%%%%%%%%%%%%%%%%%%%%%%%%%%%%%%%%%%%%

%%%%%%%%%%%%%%%%%%%%%%%%%%%%%%%%%%%%%%%%%%%%%%%%%%%%%%%%%%%%%%%%%%%%%%%%%%%%%%%%%%%%%%%%%%%%%%%%%%%%%%%%%%%%%%%%%%%%%%%%%%%%%%%%%%%%%%%%%%%%%%%%%
%%%%%%%%%%%%%%%%%%%%%%%%%%%%%%%%%%%%%%%%%%%%%%%%%%%%%%%%%%%%%%%%%%%%%%%%%%%%%%%%%%%%%%%%%%%%%%%%%%%%%%%%%%%%%%%%%%%%%%%%%%%%%%%%%%%%%%%%%%%%%%%%%

%%%%%%%%%%%%%%%%%%%%%%%%%%%%%%%%%%%%%%%%%%%%%%%%%%%%%%%%%%%%%%%%%%%%%%%%%%%%%%%%%%%%%%%%%%%%%%%%%%%%%%%%%%%%%%%%%%%%%%%%%%%%%%%%%%%%%%%%%%%%%%%%%
%%%%%%%%%%%%%%%%%%%%%%%%%%%%%%%%%%%%%%%%%%%%%%%%%%%%%%%%%%%%%%%%%%%%%%%%%%%%%%%%%%%%%%%%%%%%%%%%%%%%%%%%%%%%%%%%%%%%%%%%%%%%%%%%%%%%%%%%%%%%%%%%%

%%%%%%%%%%%%%%%%%%%%%%%%%%%%%%%%%%%%%%%%%%%%%%%%%%%%%%%%%%%%%%%%%%%%%%%%%%%%%%%%%%%%%%%%%%%%%%%%%%%%%%%%%%%%%%%%%%%%%%%%%%%%%%%%%%%%%%%%%%%%%%%%%
%%%%%%%%%%%%%%%%%%%%%%%%%%%%%%%%%%%%%%%%%%%%%%%%%%%%%%%%%%%%%%%%%%%%%%%%%%%%%%%%%%%%%%%%%%%%%%%%%%%%%%%%%%%%%%%%%%%%%%%%%%%%%%%%%%%%%%%%%%%%%%%%%

%%%%%%%%%%%%%%%%%%%%%%%%%%%%%%%%%%%%%%%%%%%%%%%%%%%%%%%%%%%%%%%%%%%%%%%%%%%%%%%%%%%%%%%%%%%%%%%%%%%%%%%%%%%%%%%%%%%%%%%%%%%%%%%%%%%%%%%%%%%%%%%%%
%%%%%%%%%%%%%%%%%%%%%%%%%%%%%%%%%%%%%%%%%%%%%%%%%%%%%%%%%%%%%%%%%%%%%%%%%%%%%%%%%%%%%%%%%%%%%%%%%%%%%%%%%%%%%%%%%%%%%%%%%%%%%%%%%%%%%%%%%%%%%%%%%

\newpage

\end{document}